\documentclass[prl,twocolumn,aps,superscriptaddress,notitlepage,floatfix,10pt]{revtex4-2}
\usepackage{amsmath,mathtools,amsthm,amssymb,pifont}
\usepackage[percent]{overpic}
\usepackage[utf8]{inputenc}
\usepackage{comment}
\usepackage[american]{babel}
\usepackage{graphicx,xcolor,bbold,titlesec}

\usepackage{MnSymbol}

\usepackage[colorlinks,
bookmarksopen,
bookmarksnumbered,
citecolor=teal,
linkcolor=teal,
urlcolor=teal]{hyperref}
\usepackage{braket}
\usepackage{enumitem}
\usepackage{subfigure}
\usepackage{ifthen}

\usepackage{orcidlink}
\usepackage{tikz,ifthen}
\usepackage{bbold}
\usepackage{tikz-network}
\usetikzlibrary{patterns,decorations.pathreplacing,calligraphy}
\usetikzlibrary{shapes,arrows.meta,decorations.pathmorphing}

\newcommand{\titleinfo}{Growth and spreading of quantum resources under random circuit dynamics}

\renewcommand{\section}[1]{\textbf{\emph{#1}}.---}

\begin{document}
\title{\titleinfo} 

\author{Sreemayee Aditya~\orcidlink{0000-0002-0412-7944}}
\email{asreemay@uni-koeln.de}
\affiliation{Institut für Theoretische Physik, Zülpicherstraße 77a, 50937, Köln, Germany}

\author{Xhek Turkeshi~\orcidlink{0000-0003-1093-3771}}
\email{xturkesh@uni-koeln.de}
\affiliation{Institut für Theoretische Physik, Zülpicherstraße 77a, 50937, Köln, Germany}

\author{Piotr Sierant~\orcidlink{0000-0001-9219-7274}}
\email{piotr.sierant@bsc.es}
\affiliation{Barcelona Supercomputing Center Plaça Eusebi Güell, 1-3 08034, Barcelona, Spain}

\begin{abstract}
Quantum many-body dynamics generate nonclassical correlations naturally described by quantum resource theories. Quantum magic resources (or nonstabilizerness) capture deviation from classically simulable stabilizer states, while coherence and fermionic non-Gaussianity measure departure from the computational basis and from fermionic Gaussian states, respectively. We track these resources in a subsystem of a one-dimensional qubit chain evolved by random brickwall circuits. 
For resource-generating gates, evolution from low-resource states exhibits a universal rise-peak-fall behavior, with the peak time scaling logarithmically with subsystem size and the resource eventually decaying as the subsystem approaches a maximally mixed state.
Circuits whose gates do not create the resource but entangle neighboring qubits, give rise to a ballistic spreading of quantum resource initially confined to a region of the initial state. Our results give a unified picture of spatiotemporal resource dynamics in local circuits and a baseline for more structured quantum many-body systems.
\end{abstract}
\maketitle

\section{Introduction} 
Non-classical features of quantum states such as entanglement~\cite{HorodeckiEntanglement2009, AmicoEntanglemennt2008}, anticoncentration~\cite{Dalzellanticoncentration2022, SierantUniversal2022, ClaeysFock2025, MagniTurkeshi2025QuantumComplexity, SauliereUniversality2025}, and the formation of unitary designs~\cite{BoulandOn2018, LamiAnticoncentration2025, FavaDesigns2025, ChengPseudoentanglement2025, IppolitiDynamical2023, ClaeysEmergentquantum2022, SchusterRandom2025} have become central diagnostics of quantum computational power beyond classical simulability~\cite{Feynman1982Simulating, Cirac12, Bharti22}. Quantum resource theories~\cite{Chitambar2019} of entanglement~\cite{HorodeckiEntanglement2009, AmicoEntanglemennt2008}, nonstabilizerness~\cite{Veitch14, LiuWintermagic2022, leone2022stabilizer, Bravyi2016magic, Wang2019magic, Howard17rom}, coherence~\cite{Baumgratzcoherence2014, Streltsovcoherence2017, Saxenacoherence2020}, and bosonic~\cite{ZhuangNG2018, TakagiNG2018} or fermionic~\cite{Hebenstreit2019all,Lumia24gaussian, Lyu24fermi, Coffman25fermi, Sierant25fermi} non-Gaussianity provide a unified framework to quantify such non-classicality, which underpins fault-tolerant quantum computation as well as metrological and thermodynamic advantages. In many-body dynamics, quantum resources track the complexity of states generated by time evolution, providing a lens on quantum chaos~\cite{Haakebook, Lauchli08, Kim13}, information scrambling~\cite{Garcia23, Swingle16, Yunger19, Vikram24, Kaneyasu25, Odavic25}, thermalization~\cite{D_Alessio_2016, Pappalardi2022ETHFreeProbability, Abanin2019MBLColloquium, Sierant2025MBLReview, tirrito2025universalspreadingnonstabilizernessquantum, Jasser25, Bera25syk, Sticlet2025nonstabilizernesstransport, Falcao25mbl}, properties of ground and critical states~\cite{haug2023quantifying, Tarabunga2023criticalitytogauge, TarabungaCastelnovo2024magicRokhsar, Collura2024NonStabFermGauss, Santra2025Latticegauge}, conformal field theories~\cite{White2021CFTemagical, Frau2025StabDisentanglingCFT, Hoshino25sre}, monitored systems~\cite{Bejan2024magictrasnition, Fux2024magictransition, Tirrito2025MagicGauss, Trigueros2025NoisyMagic}, and metrological responses~\cite{Hernandez25nonstab}.

Capturing the dynamical behavior of quantum resources in many-body settings typically requires computing nonlinear resource measures (or monotones)~\cite{CoeckeResourcetheory2016,Chitambar2019}. Random quantum circuits~\cite{fisher2023random, Nahumoperatorspreading2018, Chanquantumchaos2018, Shivamquantumchaos2023} provide minimally structured models of local quantum dynamics and have yielded a detailed understanding of entanglement growth via the minimal-membrane picture~\cite{Nahumentanglement2017, Zhou2019EmergentStatMech, Vasseur2019HolographicRTN, Sierant2023MinimalMembranes, Zhou2020EntanglementMembrane, Sommers2024ZeroTempMembranes} and information scrambling~\cite{Nahumoperatorspreading2018, vonKeyserlingk2018hydrodynamics}. Recent work has extended this analysis to coherence~\cite{Bertoni2024ShallowShadows, TurkeshiHilbert2024, GarciaMartin2024SymplecticCircuits, Braccia2024ExactMoments, Christopoulos2025UniversalOverlaps, Magni2025AnticoncentrationClifford, ZhangEtAl2025DesignsMagicAugmented} and nonstabilizerness~\cite{turkeshi2025magic}, showing that these resources saturate on timescales scaling only logarithmically with system size, much faster than the system-size-linear saturation of entanglement entropy. While ergodic Floquet models exhibit analogous phenomenology, local Hamiltonian dynamics yields a parametrically slower resource growth~\cite{Tirritoanticoncentration2025}.

Most of these works, however, focus on \emph{global} resource measures, such as the stabilizer entropy~\cite{leone2022stabilizer} or participation entropy~\cite{DeLuca13, Mace19,Stephan09, Stephan10, Luitz14universal, Luitz14improving, Atas12, Luitz14improving, Lindinger19,Pausch21, Li25PE} of the full many-body state. Global metrics quantify how much resource the system carries overall, but do not track its distribution and propagation. Fault-tolerant architectures, however, require logical operations that spread errors only locally, e.g., via geometrically local gates~\cite{EastinKnill2009transversalgate, BravyiKoenig2013localstablizer, Bombin2016jump}. Local resource spreading can also underlie the quantum Mpemba effect~\cite{turkeshi2025quantum, LiuMpemba2024, Aditya25Mbempa, summer2025resourcetheoreticalunificationmpemba}. Hence, for understanding information scrambling, thermalization, and experimentally relevant local probes, the \emph{local} structure of quantum resource growth is the key.

\begin{figure}
\centering
\begin{overpic}[width=0.48\textwidth]{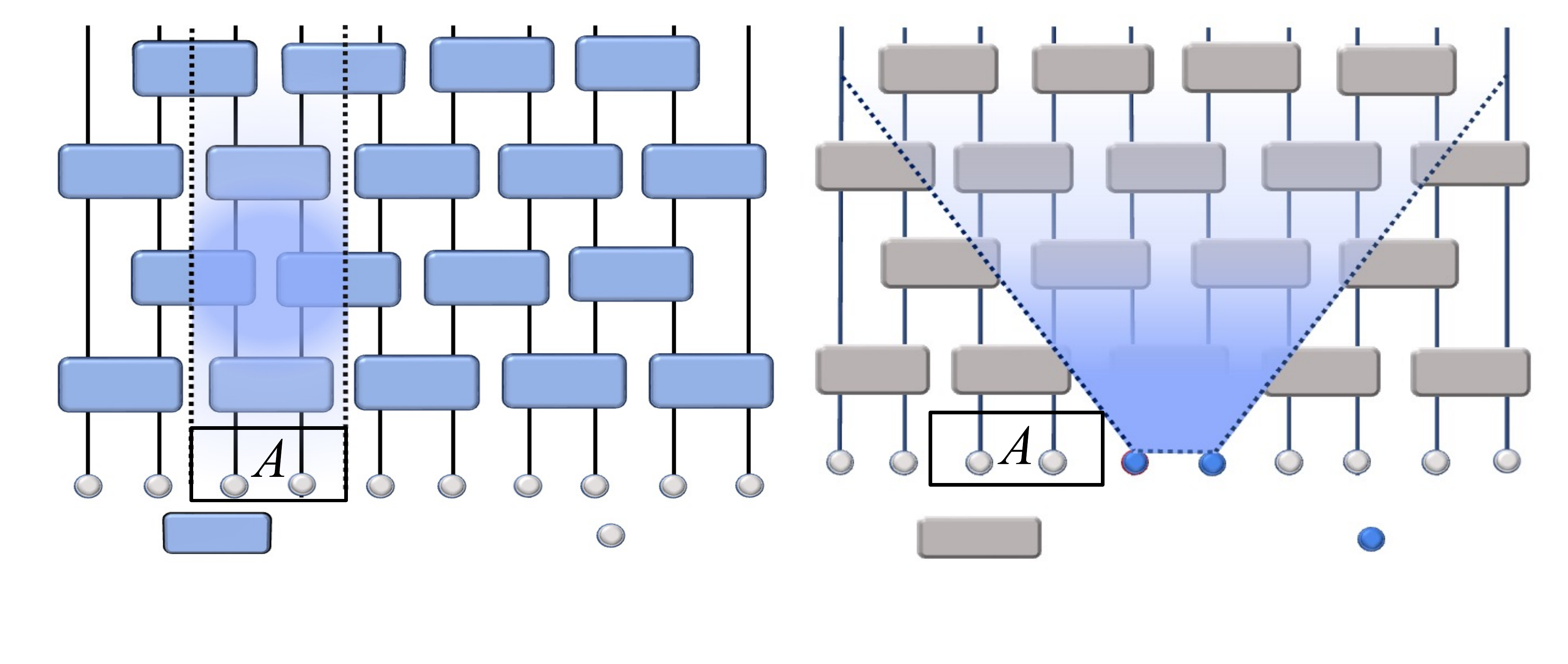}
\put(0.05,39){(\textbf{a})  }
\put(48.9,39){(\textbf{b})}  
\put( 5,-0){\shortstack{Resourceful\\ gate}  } 
\put(29,-0){\shortstack{Resource-free\\ state}}
\put(55,-0){\shortstack{Resource-\\free gate}}
\put(82,-0){\shortstack{Magic\\state}}
       
\end{overpic}
\caption{Our setup. (a) \emph{Local resource dynamics:} We track the resource content of a subsystem $A$ under brick-wall circuit dynamics with resource-generating gates starting from a resource-free initial state, uncovering a universal rise–peak–fall profile. (b) \emph{Resource spreading:} Starting from an initial state where the resource is confined to a small region and evolving with circuits whose gates do not generate the resource, we scan the spatiotemporal position of $A$ and expose ballistic spreading of the resource.}
\label{fig1:cartoon}
\end{figure}
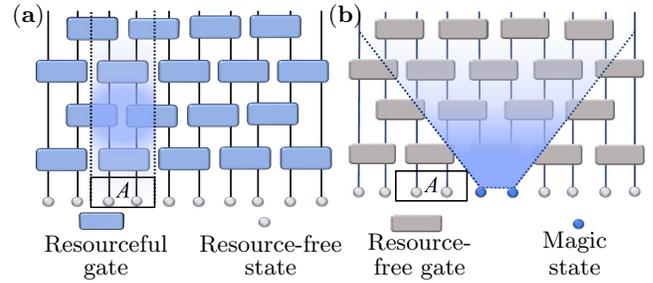

In this Letter, we study the \emph{local dynamics} of quantum resources in one-dimensional random brickwork circuits, focusing on three paradigmatic resources: nonstabilizerness, coherence, and fermionic non-Gaussianity. We consider a subsystem $A$ and analyze two complementary setups, see Fig.~\ref{fig1:cartoon}. In the first setup, circuits are composed of random, resource-generating two-qubit gates acting on an initial computational-basis product state. This allows us to characterize \emph{local resource dynamics}, from the initial buildup induced by the gates to the subsequent decay of the resource content of $A$ as it approaches a maximally mixed state. In the second setup, we consider circuits whose gates do not generate the resource, acting on an initial state where the resource is localized in a finite region. By shifting the position of $A$, we reveal ballistic spreading of the initially localized resourceful cluster.

\section{Quantum resources} Resource theories provide a general framework for quantifying nonclassical features of quantum states by distinguishing a set of \emph{free states}, which can be prepared without cost, and \emph{free operations}, which do not generate the resource from free states. Any state outside the free set is deemed resourceful. This naturally leads to definition of monotones $M(\rho)$, i.e., functions of state $\rho$ of $L$ qubits that satisfy two basic requirements: (i) \emph{faithfulness} -- $M(\rho)=0$ if and only if $\rho$ is a free state, and (ii) \emph{monotonicity} -- $M(\Lambda[\rho]) \leq M(\rho)$ for all free operations $\Lambda$. 

Stabilizer states, denoted as $\ket{\sigma}$, are pure states obtained from $\ket{0}^{\otimes L}$ by action of Clifford unitaries, i.e., operators that map a Pauli string (an $L$-fold tensor product of Pauli operators $I, X, Y, Z$) to a Pauli string. In the resource theory of nonstabilizerness the \emph{free states} belong to the convex hull $\textsf{STAB} \equiv \{ \sum_i p_i \ket{\sigma_i} \bra{\sigma_i}: p_i > 0, \, \sum_i p_i = 1 \}$ of pure stabilizer states. The \emph{free operations} are stabilizer protocols~\cite{Haug2023stabilizerentropies}, that include action of Clifford unitaries.
Stabilizer entropy~\cite{leone2024stabilizer}, is an experimentally relevant~\cite{Oliviero2022MeasuringMagic, Niroula23phase, Haug24efficient, Haug25witness} measure of nonstabilizerness, which, however, has the monotonicity property only for pure states~\cite{Haug2023stabilizerentropies, leone2024stabilizer}. Since we aim at quantifying nonstabilizerness of a subsystem $A$ consisting of $L_A$ qubits that is generally in a mixed state $\rho_A$, we utilize the log-robustness of magic (LRoM)~\cite{Howard17rom, Heinrich2019RobustnessMagic,Hamaguchi2024handbook} defined as 
\begin{equation}
  \mathcal{L}(\rho_A) = \log\min_{x_1,...,x_k}
  \left\{ \sum_{i=1}^k |x_i|: \rho_A = \sum_{i=1}^{k} x_i \sigma_i,\ \right\},
\label{eq:RoM}
\end{equation}
where $k$ is an integer. The LRoM quantifies the distance of $\rho_A$ from $\textsf{STAB}$ and is a faithful monotone of the resource theory of nonstabilizerness.

Coherence is at the heart of interference phenomena. In the theory of quantum coherence, one distinguishes a specific basis, chosen here as the computational basis $\mathcal{B} = \{ \ket{i}\}_{i=1,\ldots,2^L}$ and selects the set of \emph{free states} as the convex hull $\textsf{INCOH}=\{ \sum_i p_i |i \rangle \langle i|\,: p_i > 0, \, \sum_i p_i = 1 \}$, referred to as the incoherent states. \emph{Free operations} in the theory of coherence~\cite{Baumgratzcoherence2014, Levi14} are not able to create coherence from the incoherent states, and the relative entropy of coherence~\cite{Baumgratzcoherence2014, Levi14, Streltsovcoherence2017,Saxenacoherence2020},
\begin{equation}
  C_{d}(\rho_A) = S(\rho_A^{D}) - S(\rho_A),
  \label{eq:cd}
\end{equation}
where $S(\rho) = - \mathrm{tr}\!\left(\rho \log_2 \rho\right)$ is the von Neumann entropy and $\rho_A^{D} = \sum_{z} \langle z | \rho_A | z \rangle \, | z \rangle \langle z |$, is a faithful monotone of coherence of the subsystem $A$.

Fermionic Gaussian states are of the form $\rho_G = c\exp(\frac{i}{2}\sum_{mn} g_{mn} \gamma_m\gamma_n) $~\cite{Mbeng24,Surace22}, where $\gamma_{2k-1} = \left(\prod_{m=1}^{k-1} Z_m\right)X_k$, $\gamma_{2k} = \left(\prod_{m=1}^{k-1} Z_m\right)Y_k$ are known as Majorana operators, and coefficients $g_{mn}$ form an antisymmetric $2L \times 2L$ matrix while $c$ is a constant. In the resource theory of fermionic non-Gaussianity $\textsf{GAUSS} = \{ \rho_G \} $ is the set of \emph{free states}, while \emph{free operations} include fermionic Gaussian unitaries, i.e. operators of the form $U_G = \exp(\frac{1}{2} \sum_{mn}h_{mn}\gamma_n \gamma_m)$. A faithful monotone of fermionic non-Guassianity is the relative entropy of non-Gaussianity~\cite{Lumia24gaussian, Lyu24fermi}
\begin{equation}
  \mathcal{NG}(\rho_A) = S(\rho^{F}_A) - S(\rho_A),
  \label{eq:rel_entropy_gaussianity}
\end{equation}
where $\rho^F_A$ is a fermionic Gaussian state with the covariance matrix $\Gamma_{ab}= -\frac{i}{2}\,\mathrm{tr}\!\left[\rho_A (\gamma_a\gamma_b-\gamma_b\gamma_a )\right]$. The von Neumann entropy of $\rho^{F}_A$ is evaluated with the techniques for free-fermions~\cite{Peschel09} as $S(\rho^F_A) = \sum_{j=1}^{L_A}H\!\left(\frac{1 + \lambda_j}{2}\right)$, where $\lambda_j$ are the Williamson eigenvalues~\cite{Williamson36, Zumino62} of the covariance matrix $\Gamma_{ab}$ and  $H(x) = - x \log_2 x - (1 - x)\log_2(1 - x)$.

\section{Random circuits} 
We consider a one-dimensional chain of $L$ qubits,  see Fig. \ref{fig1:cartoon}, evolving under brick-wall circuit. The evolution operator reads $U_t = \prod_{r=1}^t U^{(r)}$, where $t$ is the circuit depth, referred to as time. The layers $U^{(r)}$ of the circuits are fixed as
\begin{equation}
    U^{(2m)} = \prod_{i=1}^{N/2-1} u_{2i,2i+1}\;,\quad U^{(2m+1)} = \prod_{i=1}^{N/2} u_{2i-1,2i}\;,
    \label{eq:randC}
\end{equation}
where each two-qubit gate \(u_{i,j}\) is drawn with probability \(\epsilon\) from an ensemble (depending on the setup and resource) and is the identity \(\mathbb{1}_{i,j}\) with probability \(1-\epsilon\). The dilution parameter \(\epsilon\) controls the rate of the circuit dynamics.

In the first setup, we fix the subsystem $A$ and starting from the computational basis state $\ket{\psi_1} = \ket{0\cdots 0}$, we monitor the resource monotones \eqref{eq:RoM}-\eqref{eq:rel_entropy_gaussianity} in the reduced state of the subsystem $\rho_A(t)=\mathrm{tr}_{\bar A}\bigl[\ket{\psi(t)}\bra{\psi(t)}\bigr]$, where  $\ket{\psi(t)} = U_t \ket{\psi_1}$ is the time-evolved state and $\bar{A}$ denotes the complement of $A$. For suitable gate ensembles, this setup enables us to track how the circuits generate the resources acting on $\ket{\psi_1}$ which is a free state of for theories of nonstabilizerness, coherence and non-Gaussianity.

In the second setup, the circuit comprises gates that are free operations, while the resource is localized in a finite ``cluster'' of the initial state $\ket{\psi_2}$, taken of the form $\ket{\psi_2} = \ket{0\ldots0}\otimes \ket{m} \otimes \ket{0 \ldots 0}$, where $\ket{m}$ denotes a resourceful state localized on $L_M$ central sites of the chain. This setup enables us, by varying the position of the subsystem $A$, to track how the resources spread throughout the chain under the dynamics of the circuits.

\begin{figure*}
\centering
\includegraphics[width=0.95\textwidth]{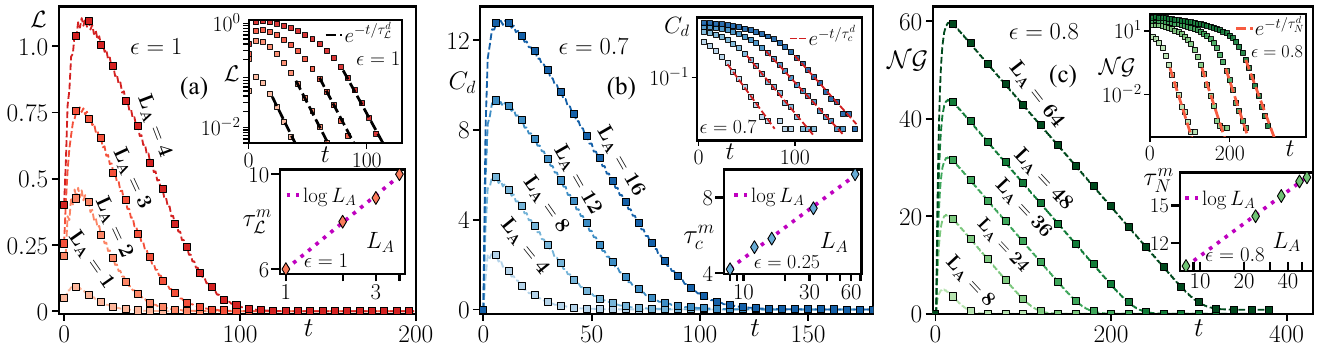}
\caption{
Local dynamics of (a) LRoM \(\mathcal{L}(\rho_A)\) (\(L=24\)), (b) relative entropy of coherence \(C_d(\rho_A)\) (\(L=128\)), and (c) relative entropy of non-Gaussianity \(\mathcal{NG}(\rho_A)\) (\(L=128\)) for subsystems of size \(L_A\) under resource-generating circuits initialized in the free state \(\ket{\Psi_1}\). Results are averaged over at least \(10^4\) circuit realizations; the dilution parameter \(\epsilon\) in each panel is chosen for clarity. Main panels: the rise-peak-fall structure of resource content of $\rho_A(t)$. Top insets: exponential relaxation of the monotones at large $t$. Bottom insets: peak times scale logarithmically with subsystem size, \(\tau^{m} \sim \log L_A\).
}
\label{fig2}
\end{figure*}

\section{Local resource growth and decay} Now, we study the first setup by fixing the initial state as $\ket{\psi_1}$ and monitoring the resource content of the subsystem state $\rho_A$.

\paragraph{Nonstabilizerness.} 
To track the dynamics of LRoM, \eqref{eq:RoM}, we choose the gates \(u_{i,j}\) in~\eqref{eq:randC} as Haar-random unitaries from the unitary group \(\mathcal{U}(4)\). Fixing the system size to \(L = 24\) and considering \(L_A \leq 4\), we perform full statevector simulation to compute \(\rho_A(t)\) and then we evaluate \(\mathcal{L}(\rho_A(t))\)~\cite{Hamaguchi2024handbook}. The results are shown in Fig.~\ref{fig2}(a). The gates generate nonstabilizerness and \(\mathcal{L}(\rho_A)\) grows, reaching a maximum at time \(\tau_{\mathcal{L}}^{m}\). This maximum occurs at a time \(\tau_{\mathcal{L}}^{m} \sim \log L_A\), as shown in the bottom inset of Fig.~\ref{fig2}(a). The logarithmic scaling of \(\tau_{\mathcal{L}}^{m}\) is reminiscent of the logarithmic-in-system-size saturation of global nonstabilizerness measures under random circuit dynamics~\cite{turkeshi2025magic}. At later times, the dynamics strongly entangles subsystem \(A\) with its complement \(\bar{A}\), and \(\rho_A\) gradually approaches the maximally mixed state, \(\rho_A \stackrel{t \gg 1}{\mapsto} \mathbb{1}_{A}/2^{L_A} \), where \(\mathbb{1}_A\) is the identity operator on \(A\). The maximally mixed state belongs to \(\textsf{STAB}\), and LRoM decays exponentially to $0$, \(\mathcal{L}(\rho_A(t)) \propto e^{-t/\tau^d_\mathcal{L}}\), with a decay time \(\tau^d_{\mathcal{L}}\) independent of the subsystem size; \(\mathcal{L}(\rho_A(t))\) reaches a fixed threshold \(\theta \ll 1\) on a timescale proportional to \(L_A\), see~\cite{supmat} for further analysis.

\paragraph{Coherence and non-Gaussianity.}
The characteristic rise-peak-fall structure of the local resource content under local random circuit dynamics is also found for the measures of coherence~\eqref{eq:cd} and fermionic non-Gaussianity~\eqref{eq:rel_entropy_gaussianity}, as we demonstrate in the following. To obtain a clearer picture of the local resource dynamics, we restrict the gates \(u_{i,j}\) in~\eqref{eq:randC} to be Clifford gates, which enables numerical simulations of systems comprising hundreds of qubits via the tableaux formalism~\cite{AaronsonGottesman2004}, far beyond the reach of full statevector simulations.

To study the relative entropy of coherence of \(\rho_A\), we choose the gates \(u_{i,j}\) from the subset \(C_2 \setminus C_2^{\mathrm{inc}}\) of the \(11\,520\) two-qubit Clifford gates \(C_2\), where \(C_2^{\mathrm{inc}}\) denotes the set of \(768\) \emph{incoherent} Clifford unitaries that do not generate coherence (superpositions of two or more computational-basis states) when acting on computational-basis states. Using numerical simulations~\cite{Gidney2021stim} of a chain of \(L = 128\) qubits and computing \(C_d(\rho_A)\) within the tableaux formalism via the formula derived in~\cite{Aditya25Mbempa}, we obtain the results shown in Fig.~\ref{fig2}(b). As in the case of nonstabilizerness, we observe a rise-peak-fall structure of \(C_d(\rho_A(t))\), with a peak at times \(\tau^m_c \propto \log L_A\), reminiscent of the log-depth anticoncentration~\cite{Dalzellanticoncentration2022, TurkeshiHilbert2024}, now visible over a considerably wider range of subsystem sizes \(L_A \leq 64\), and an exponential decay \(C_d(\rho_A(t)) \propto e^{-t/\tau^d_c}\) for \(t \gg 1\). The dynamics saturates down to a fixed threshold \(\theta \ll 1\) on timescales scaling linearly with \(L_A\), see also~\cite{supmat}.

Clifford gates \(C_2\) contain a subset of \(192\) Clifford matchgates \(C_2^{\mathrm{Gauss}}\), which are Clifford unitaries that also fulfill the matchgate condition~\cite{Valiant01, Terhal02, Jozsa08}, i.e., they are fermionic Gaussian unitaries. To study the evolution of \(\mathcal{NG}(\rho_A(t))\), we choose the gates \(u_{i,j}\) in~\eqref{eq:randC} randomly from the set \(C_2 \setminus C_2^{\mathrm{Gauss}}\), and simulate a chain of \(L = 128\) qubits within the tableaux formalism~\cite{Gidney2021stim}. The results shown in Fig.~\ref{fig2}(c) again exhibit a rise--peak--fall structure, with the peak time \(\tau^m_N \propto \log L_A\),
consistently with the similarity of fermionic magic resources growth~\cite{Sierant25fermi} to dynamics of coherence and nonstabilizerness, and an exponential decay \(\mathcal{NG}(\rho_A(t)) \propto e^{-t/\tau^d_N}\) at long times.

\begin{figure*}
\centering
\includegraphics[width=0.335\textwidth]{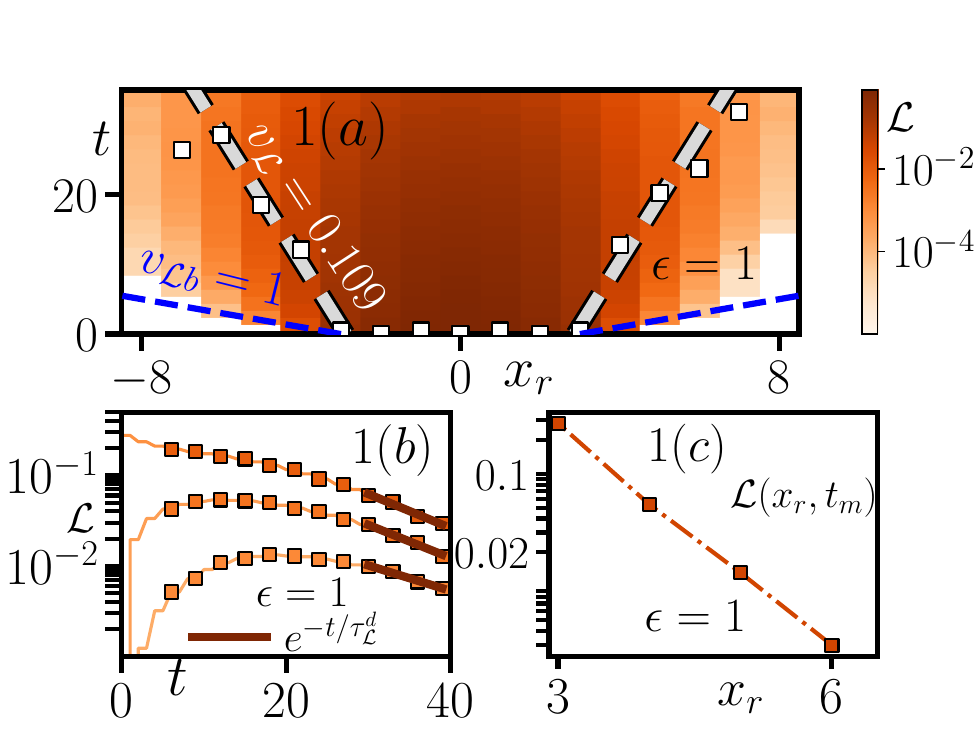}%
\includegraphics[width=0.335\textwidth]{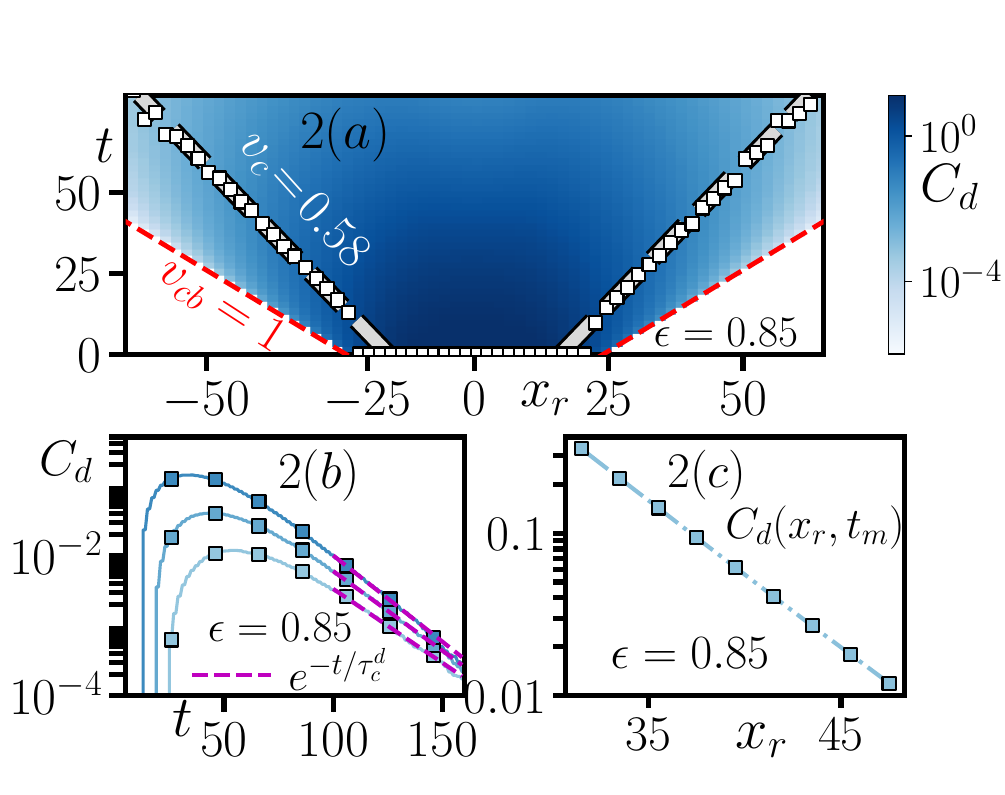}%
\includegraphics[width=0.335\textwidth,height=4.75cm]{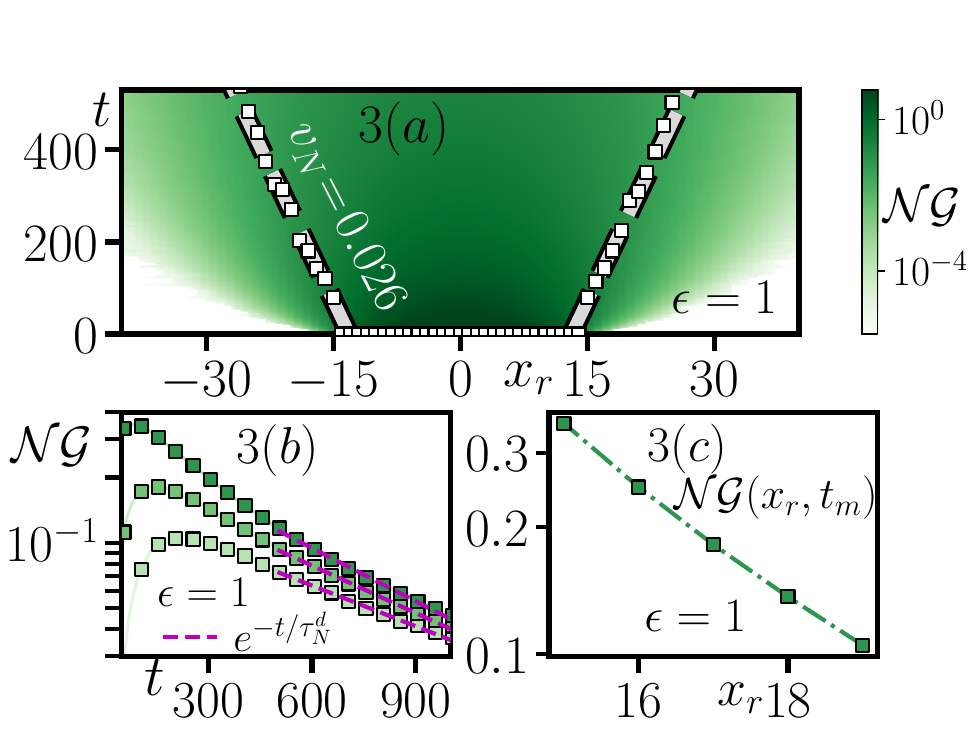}
\caption{
Spreading of (1) nonstabilizerness ($L=24$), (2) coherence ($L=256$), and (3) non-Gaussianity ($L=128$) from a localized resource cluster under circuits comprising resource-free gates.
The results are averaged over $10^4$ circuit realizations. Panels (a) display the emergence of ballistic light cone structure of the spreading; panels (b) show the evolution of the resource content of $\rho_A$ at fixed $x_r$ ($-x_r=(5,4,3), (47.5,41.5,35.5)$ and $(19,17,15)$ for $R=\mathcal{L},c, N$, respectively.); panels (c) highlight the exponential attenuation of the peak local resource with distance along the light cone $x = v_R t + \mathrm{const}$ (with $R=\mathcal{L}, c, N$).}
\label{fig3}
\end{figure*}

\section{Resource spreading}
We now turn to the second setup, where the system is initialized in a locally resourceful state \(\ket{\psi_2}\) and the gates \(u_{i,j}\) are chosen as free operations of the respective resource theory.

\paragraph{Nonstabilizerness.}To study the spreading of nonstabilizerness, we draw each gate \(u_{i,j}\) uniformly from the set \(C_2\) of two-qubit Clifford gates, and prepare the initial state \(\ket{\psi_2} = \ket{0\ldots0}\otimes \ket{m} \otimes \ket{0 \ldots 0}\), with \(\ket{m} = \ket{T}^{\otimes L_M}\) a product of single-qubit magic states \(\ket{T} = \cos(\pi/8)\ket{0} + \sin(\pi/8)\ket{1}\)~\cite{bravyi2005universal}. We fix \(L_M = 4\), perform statevector simulations for \(L = 24\) qubits to obtain the time-evolved state \(\ket{\psi(t)} = U_t\ket{\psi_2}\), and compute \(\mathcal{L}(\rho_A(t))\). We introduce a variable \(x_r \in [-L/2, L/2)\) denoting the relative position of the center-of-mass of subsystem \(A\) respective to the center of the magic region. The resulting spatiotemporal profile of \(\mathcal{L}(\rho_A)\) is shown in Fig.~\ref{fig3}(1a).

The spacetime profile of nonstabilizerness exhibits a clear light-cone structure, signaling ballistic spreading of the resource. The earliest detectable nonstabilizerness time grows linearly with the distance \(x_r\) of subsystem \(A\) from the initial magic cluster, defining an outer light cone expanding with velocity $v_{\mathcal{L}b} = 1$ set by gate locality. Along this front, \(\mathcal{L}(\rho_A)\) decays exponentially with distance and at larger distances (e.g. $x_r=8$ and $t=8$), it falls below the threshold \(10^{-6}\). After the first nonzero signal at fixed \(x_r\), the nonstabilizerness of \(\rho_A\) increases, see Fig.~\ref{fig3}(1b), reaching a maximum at times $t_m$ satisfying \(x_r = v_{\mathcal{L}} t_m + \mathrm{const}\), which reflects ballistic spreading of nonstabilizerness. The peak LRoM decays exponentially with \(x_r\) [Fig.~\ref{fig3}(1c)]. At late times, \(\mathcal{L}(\rho_A)\) decays exponentially back to zero [Fig.~\ref{fig3}(1b)] as \(\rho_A\) approaches the maximally mixed state. This return of every small subsystem to a free state at long times shows that magic is not a conserved local density under free two-body dynamics: the initially localized resource is transferred into highly nonlocal correlations, so that local magic quantified by \(\mathcal{L}(\rho_A)\) vanishes even though global nonstabilizerness measures of \(\ket{\psi(t)}\) remain constant at all \(t\), since \(U_t\) consists of stabilizer operations.

\paragraph{Coherence and non-Gaussianity.} We restrict the gates \(u_{i,j}\) to Clifford unitaries~\cite{AaronsonGottesman2004, Gidney2021stim} to study the spreading of coherence and fermionic non-Gaussianity in chains comprising hundreds of qubits.

For coherence, we choose each \(u_{i,j}\) to be the SWAP gate with probability \(p\), and with probability \(1-p\) a randomly chosen gate from the set \(C^{\mathrm{inc}}_2\). The initial state is prepared as \(\ket{\psi_2}=\ket{0\ldots0}\otimes \ket{m} \otimes \ket{0 \ldots 0}\), with \(\ket{m} = [(\ket{0}+\ket{1})/\sqrt{2}]^{\otimes L_c}\) acting as a source of coherence. Fixing $L=256$, $L_c=32$ and $L_A=32$, we compute the relative entropy of coherence \(C_d(\rho_A)\) as a function of time \(t\) and the position \(x_r\) of subsystem \(A\). Our results, shown in Fig.~\ref{fig3}(2a)--(2c), uncover a spatiotemporal pattern of \(C_d(\rho_A)\) analogous to the nonstabilizerness case, with light cones at velocity \(v_{cb}=1\) delimiting the region of nonzero coherence and a maximum of \(C_d(\rho_A)\) propagating along a light cone with velocity \(v_c<1\).

For non-Gaussianity, we choose the gates \(u_{i,j}\) in~\eqref{eq:randC} randomly from the Clifford matchgate set \(C^{\mathrm{Gauss}}_2\), while the initial state is \(\ket{\psi_2}=\ket{0\ldots0}\otimes \ket{m} \otimes \ket{0 \ldots 0}\), with \(\ket{m} = [\tfrac{1}{4}\bigl(\ket{0000} + \ket{1100} + \ket{0011} - \ket{1111}\bigr)]^{\otimes L_N}\) a product of four-qubit fermionic magic states~\cite{Oszmaniec22}. Setting \(L=128\), \(L_N=4\), and \(L_A=16\), we obtain the results for \(\mathcal{NG}(\rho_A)\) shown in Fig.~\ref{fig3}(3a)--(3c), which reveal a spatiotemporal structure of non-Gaussianity spreading fully analogous to the nonstabilizerness and coherence cases.

\section{Discussion}
The rise-peak-fall structure of the local resource content summarized in Fig.~\ref{fig2} stems from a competition between resource injection by resource-generating gates, which increases the local resource, and the buildup of entanglement and long-range correlations at later times, which drive \(\rho_A\) towards the maximally mixed state and suppress the resource measures. A similar behavior was observed~\cite{Ebner25barrier} for entanglement flatness~\cite{Tirrito24flatness}, which bounds nonlocal nonstabilizerness~\cite{Cao25Grav}. The ballistic spreading of resources in Fig.~\ref{fig3} arises from the generation of nonlocal correlations across the chain by the circuit dynamics, which simultaneously causes an exponential attenuation of the local resource content, sharply distinguishing resource spreading from the transport of conserved local densities. To corroborate the generality of our findings, we show in the End Matter that the same phenomenology governs the spatiotemporal dynamics of mana, a nonstabilizerness monotone~\cite{Veitch14}, in qutrit chains evolved under local circuits. We further show that coherence spreading in non-Clifford circuits exhibits the same behavior, using large-scale simulations whose cost scales exponentially only in the size of the coherent cluster \(L_c\) and polynomially in the total system size \(L\), analogous to simulation methods for Clifford circuits doped with nonstabilizer gates~\cite{Bravyi2016PRL, Bravyi19simul}.

\section{Conclusions} 
In this work, we studied the spreading of quantum resources that underlie the classical complexity of quantum states: nonstabilizerness, coherence, and fermionic non-Gaussianity, in local random circuit dynamics. For resource-generating gates, we uncovered a rise--peak--decay behavior of the local resource content, with peak times scaling logarithmically with the subsystem size, \(\tau^{m} \sim \log L_{A}\). This signals a rapid onset of local resourcefulness and the buildup of nonclassical features in local subsystems on a timescale parametrically shorter than the linear timescale required to generate correlations across the subsystem. In a complementary setup, where the gates are free operations that do not generate the resource and the chain is initialized with a localized resourceful cluster, we found ballistic spreading of the local resource content. Taken together, our results demonstrate a unified phenomenology for the spatiotemporal dynamics of nonstabilizerness, coherence, and fermionic non-Gaussianity, pointing to a form of universality in resource spreading governed primarily by locality and unitarity, rather than microscopic details of the dynamics or of the specific resource theory. We therefore conjecture that the phenomenology uncovered here applies broadly to local ergodic quantum many-body systems.

Our results open several conceptual avenues for exploration. A central challenge is to develop an analytical framework for the phenomenology uncovered here, for instance by relating nonstabilizerness and related resources to the slow hydrodynamic modes~\cite{Khemani2018operatorspreading,Rakovszky2019hydrodynamics,vonKeyserlingk2018hydrodynamics,Michailidis2024hydrodynamics} of many-body dynamics or utilizing the replica trick to perform calculations, a first step in that direction in the limit of infinite local Hilbert space dimension was already done in~\cite{ZhangGu2024QuantumMagic}. It would also be compelling to move beyond random circuits to fully ergodic Floquet drives~\cite{Moessner2017Floquetmatter,Tirritoanticoncentration2025} and genuinely chaotic Hamiltonian evolutions~\cite{polkovnikov2011colloquium}, where similar universal structures may emerge in a less coarse-grained setting. Moreover, incorporating symmetries~\cite{Tirritoanticoncentration2025} and probing the role of integrability~\cite{VidmarRigol2016Gibbsensemble} and ergodicity breaking~\cite{Abanin2019MBLColloquium, Sierant2025MBLReview, Serbyn2021ergodicitybreaking, Chandran23, Turkeshi2025paulispectrum} could shed further light on more exotic regimes of quantum resource spreading. We leave these questions for future work.

\section{Acknowledgments} 
This work is dedicated to Ada.
We thank Diptiman Sen and Sumilan Banerjee for comments and suggestions, and Emanuele Tirrito for insightful discussions and collaboration on related projects.
S.A. acknowledges support from the Alexander von Humboldt Foundation as a Humboldt postdoctoral fellow.
X.T. acknowledges support from DFG under Germany's Excellence Strategy – Cluster of Excellence Matter and Light for Quantum Computing (ML4Q) EXC 2004/1 – 390534769, and DFG Collaborative Research Center (CRC) 183 Project No. 277101999 - project B01.
P.S. acknowledges fellowship within the “Generación D” initiative, Red.es, Ministerio para la Transformación Digital y de la Función Pública, for talent attraction (C005/24-ED CV1), funded by the European Union NextGenerationEU funds, through PRTR.

{\it Note Added:} While completing this manuscript, we became aware of two independent studies by Maity, Hamazaki~\cite{maity2025localspreadingstabilizerrenyi}, and Bejan, Claeys, Yao~\cite{bejan2025magicspreadingunitaryclifford} on the spreading of local magic under Clifford dynamics. While approach of our study is markedly different, the results agree in the overlapping regime.

\begin{center}
\bf{ End matter}
\end{center}

\section{Local mana spreading in qutrits} We further substantiate the picture of nonstabilizerness spreading established for qubits in the main text by examining local resource dynamics in a one-dimensional chain of $L$ qutrits (onsite Hilbert space dimension $d=3$). As a probe of the resource content in a subsystem $A$, we employ the mana~\cite{Veitch14}, a Wigner-function–based quantifier of nonstabilizerness that constitutes a faithful magic monotone for mixed states. 
Let $\omega = e^{2\pi i/d}$ and introduce the generalized Pauli operators
\begin{equation}
    Z = \sum_{m=0}^{d-1} \omega^{m} \ket{m}\!\bra{m}, 
    \qquad
    X = \sum_{m=0}^{d-1} \ket{m+1 \bmod d}\!\bra{m}.
\end{equation}
On an $L$-site chain, we denote by $X_\ell$ and $Z_\ell$ the corresponding operators acting on site $\ell$. For each phase-space point 
$r = (r_1^x,r_1^z,\dots,r_L^x,r_L^z) \in \mathbb{Z}_d^{2L}$,
we define the Pauli string $P_r = \prod_{\ell=1}^L 
    \omega^{\frac{(d+1)}{2} r_\ell^x r_\ell^z}\,
    X_\ell^{\,r_\ell^x} Z_\ell^{\,r_\ell^z}$.
These operators generate the phase-point operators~\cite{Gross2006Hudson}
\begin{equation}
    A_0 = \frac{1}{d^L} \sum_{r \in \mathbb{Z}_d^{2L}} P_r,
    \qquad
    A_r = P_r A_0 P_r^\dagger,
\end{equation}
which in turn define the discrete Wigner function
\begin{equation}
    W_r(\rho) = \mathrm{tr}\!\left(A_r \rho\right)/d^L.
\end{equation}
The nonstabilizerness of a state $\rho$ is then quantified by the mana~\cite{veitch2012negative,Veitch14, Tarabunga2024criticalbehaviorsof,turkeshi2025magic}
\begin{equation}
    \mathcal{M}(\rho) = \log_2 \left( \sum_{r \in \mathbb{Z}_d^{2L}} \bigl| W_r(\rho) \bigr| \right),
    \label{eq:mana_def}
\end{equation}
which vanishes for stabilizer states and increases monotonically with the negativity of the Wigner quasi-probability distribution.

\begin{figure}
\includegraphics[width=0.48\textwidth]{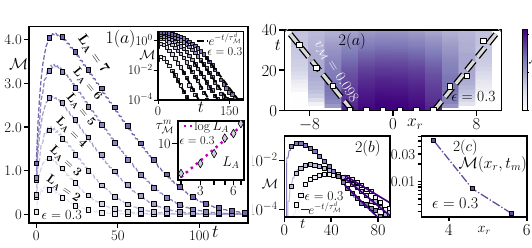}%
\caption{Dynamics of mana $\mathcal{M}(\rho_A)$ in qutrit chains. Panel (1): evolution from a resource-free initial state under resourceful two-qutrit Haar gates for $L=16$ and various $L_A$, showing a universal rise-peak-fall profile (main), late-time exponential decay (top inset), and peak times scaling logarithmically with $L_A$ (bottom inset). Panel (2): $\mathcal{M}(\rho_A)$ for $L=14$ with an initial magical cluster of size $L_M=5$ and $L_A=3$ under resource-free gates, 
exhibiting a light-cone spreading pattern and late-time exponential decay to a free state. Results averaged over at least $10^4$ circuit realizations and overall dilution $\epsilon$ in each panel are chosen for clarity.}
\label{fig4}
\end{figure}

With these qutrit degrees of freedom in place, we perform an analysis fully analogous to the qubit case. In the first setup, starting from the resource-free product state $\ket{0\cdots0}$ and evolving under local two-qutrit Haar-random gates—we find that the local nonstabilizerness in $A$ displays the same universal rise–peak–fall profile as the one observed for qubits, illustrated in Fig.~\ref{fig4}(1a)–(1c) for $L=16$ and $L_A=2$–$7$.
In particular, the timescale $\tau_M^m$ at which the reduced mana in $A$ attains its maximum again grows only logarithmically with the subsystem size, $\tau_{M}^{m}\propto \log L_A$ [bottom inset of Fig.~\ref{fig4}(1c)], thereby reinforcing the fast-scrambling phenomenology of local magic spreading across different on-site Hilbert space dimensions. Furthermore, Fig.~\ref{fig4}(2a)–(2c) shows the space–time pattern of local mana in a chain of $L=14$ qutrits, initialized in a state featuring a magic cluster of size $L_M=5$ where each single-qutrit magic state being $\ket{m}=e^{-i\frac{2\pi}{9}(X_{i}+X^{\dagger}_{i})}\ket{0}$ and evolved under a global brickwork circuit composed of random uniformly sampled two-qutrit Clifford gates $u\in C_{2}(d)$~\cite{Gottesman99solitons}. The resulting mana fronts for the subsystem of size $L_A=3$ closely parallel the nonstabilizerness spreading observed in the qubit case (Fig.~\ref{fig2}(1a)–(1c)), indicating that the qualitative features of local magic transport are robust to the underlying on-site dimensionality.

\section{Local coherence spreading using sparse-vector simulation}
To corroborate our findings in the coherence case, we additionally employ sparse-vector simulations, which enable access to system sizes far beyond the reach of 
the full statevector simulations. The central structural property we exploit is that the coherence-preserving local two-qubit gates 
are of the form $U_{\pi,\vec{\phi}} = \sum_{b \in \{00,01,10,11\}} e^{i \phi_b} \ket{\pi(b)}\bra{b}$, where $\pi$ is one of the $24$ permutations of the two-qubit computational-basis states and $\vec{\phi}\equiv (\phi_b)$ denotes four arbitrary phases, so that the gate merely permutes computational-basis states and decorates them with phases.  Consider a one-dimensional qubit chain of length $L$ with an
prepared in a state $ \ket{\psi_2}=\ket{0\ldots0}\otimes \ket{m} \otimes \ket{0 \ldots 0}$, where $\ket{m}$ is a superposition of basis states on $L_c$ qubits. The state $\ket{\psi_2}$ is supported on exactly $N_{+}(0) = 2^{L_c}$ computational-basis configurations out of the full Hilbert space of dimension $2^{L}$. Because the global time-evolution operator is built as a product of 
two-qubit gates that only permute the local states and add phases, the number of nonzero amplitudes is conserved at all times, $N_{+}(t) = N_{+}(0) = 2^{L_c}$ for every circuit layer $t$. This conservation of sparsity of the statevector is what renders the sparse-vector simulation efficient even for large $L$: although the underlying Hilbert space is exponentially large, the dynamics is confined to a smaller subspace of size $2^{L_c} \ll 2^{L}$, and it suffices to keep track of only 
$2^{L_c}$ nonzero amplitudes to compute the state of the system at any circuit depth $t$. This yields a computational cost scaling exponentially with $L_c$ but only polynomially with system size $L$, similarly to the approach of~\cite{Bravyi2016PRL, Bravyi19simul} to Clifford circuits doped with beyond Clifford gates.

\begin{figure}
    \centering
    \includegraphics[width=0.4\textwidth]{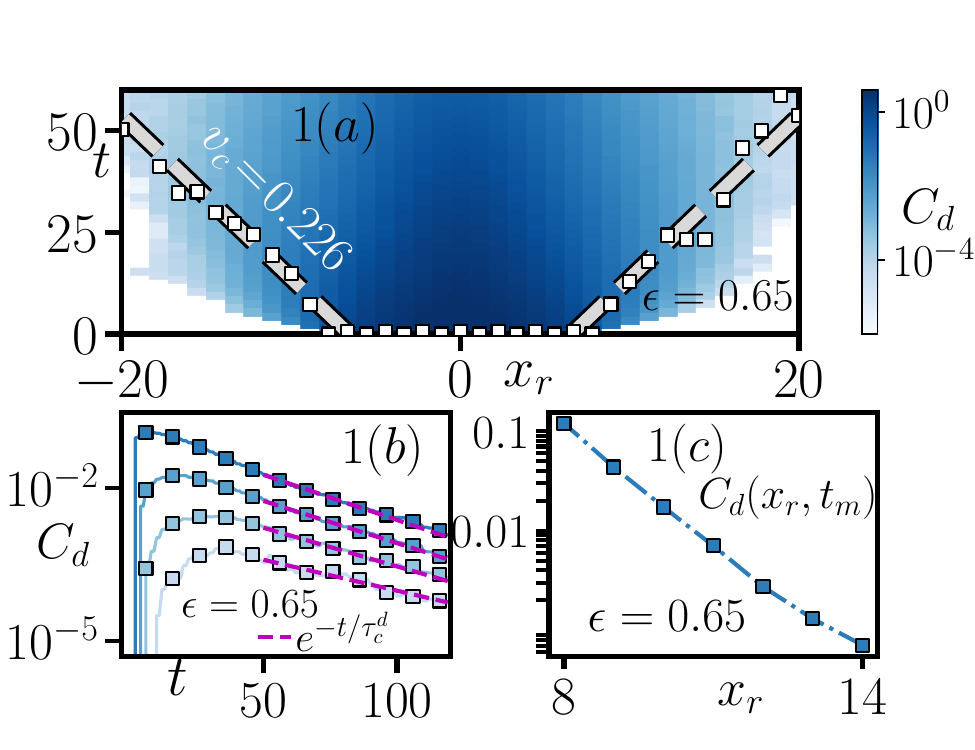}
    \caption{
    Coherence spreading for a chain of $L = 64$ qubits from initially localized resource cluster, evolved under random coherence preserving gates and computed via sparse-vector simulations, with $L_c=0$ and $L_A = 8$.
    Panel (a) reveals a clear two-front structure in the dynamics, with a sharply defined inner light cone. Panel (b-c) illustrates the late-time relaxation of the local resource content back to a resource-free steady state and exponential attenuation of the peak local resource with distance from the initial cluster.
}
    \label{fig:cohsparse}
\end{figure}

In our numerical simulations, we investigate a chain of total size $L = 64$ with an initial magic cluster of size $L_c = 8$ prepared in $\ket{+}$ states and embedded in a background of resource-free states $\ket{0}$, such that $N_{+}(t) = 2^{L_c} = 256$ for all $t$. The dynamics is generated by the
coherence-preserving two-qubit gates specified above, the 
relative entropy of coherence $C_d(\rho_A)$ is evaluated for a subsystem of size $L_A = 8$, the overall gate dilution is fixed at $\epsilon = 0.65$, and all observables are averaged over $10^{4}$ random circuit realizations. Within this approach, we obtain results shown in Fig.~\ref{fig:cohsparse} finding behavior that is quantitatively similar to the Clifford case analyzed in Fig.~\ref{fig3}(2a)-(2c). In particular, we observe a clear ballistic spreading of the local resource content $C_d(\rho_A)$ and an exponential suppression of $C_d(\rho_A)$ at larger times and distances.

\section{Ballistic propagation of outer front in the non-Gaussianity spreading}
In Fig.~\ref{fig3}, we observed that the spatiotemporal region of nonzero nonstabilizerness and coherence is delimited by a light cone corresponding velocity $v_{\mathcal{L}b}=1=v_{cb}$. In contrast, a light cone delimiting the region of non-vanishing $\mathcal{NG}(\rho_A)$ corresponds to a smaller velocity.
To elucidate this behvaior,
we uncover a directional structure within the set of Clifford matchgates: there exist two distinguished classes matchgates, one transporting the resource of non-Gaussianity ballistically to the left and the other ballistically to the right. Crucially, no single matchgate generates simultaneous left- and right-moving ballistic propagation. Moreover, the associated ballistic velocity is comparatively smaller, 
resulting in non-Gaussianity spreading still ballistically, but slower than the coherence and nonstabilizerness. 
In addition, the resource content in reduced subsystems away from the magic center decays exponentially with distance, further suppressing long-range non-Gaussian correlations. Taken together, these effects explain why a sharp ballistic outer front is not readily visible 
for the non-Gaussianity spreading.

\begin{figure}[htb]
\includegraphics[width=0.45\textwidth]{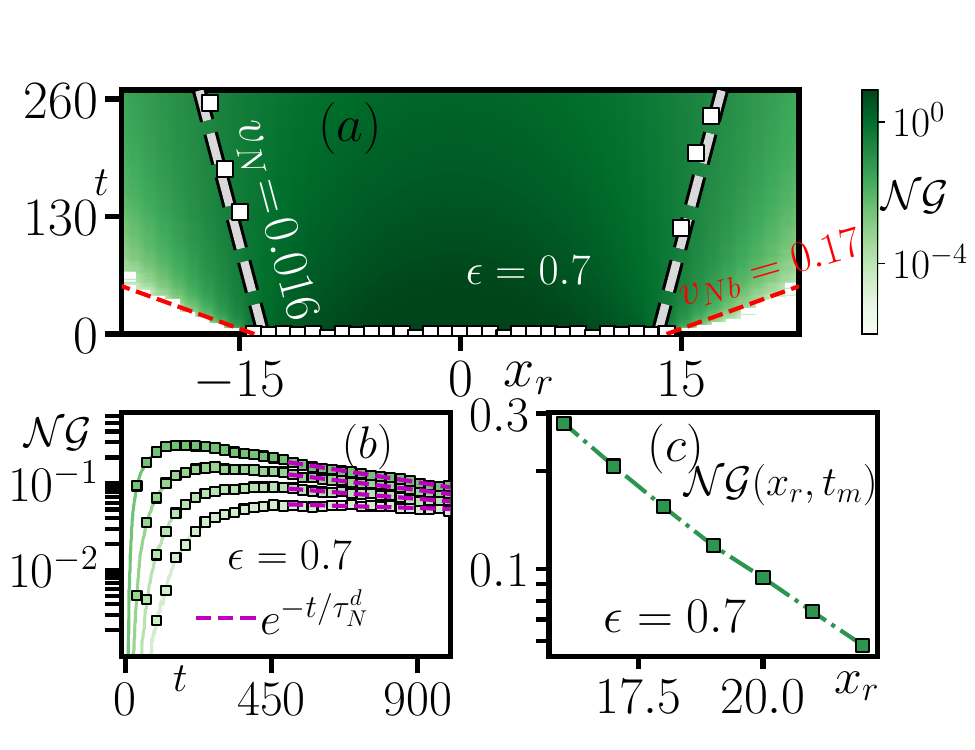}
\caption{Spatiotemporal spreading of a localized non-Gaussian magic cluster in a 128-qubit chain under resource-free two-body Gaussian (matchgate) dynamics, selecting $u_{i,j}$ gates  from $10$ Clifford matchgates propagating the initial non-Gaussian cluster in left- and right-moving ballistic modes and $10$ generic Clifford matchgates. We fix $(L_M,L_A)=(16,16)$, set $\epsilon=0.7$, and average over $10^4$ circuit realizations. The outer front in panel (a) is markedly more visible than in the case of Fig.~\ref{fig3}(c), highlighting the role of the left/right propagating gates; all other features in panels (a--c) remain qualitatively the same as in Fig.~\ref{fig3}.
}
\label{fig6}
\end{figure}

To sharpen this picture, we perform numerical simulations in which we explicitly include both the left- and right-moving Clifford matchgates, supplemented by an additional set of ten generic matchgates, for a chain of length $L = 128$
and averaging the results over at least $10^{4}$ circuit realizations. In this tailored setting, the ballistic propagation of the region of nonzero $\mathcal{NG}(\rho_A)$ 
becomes strikingly more pronounced than in the fully generic matchgate ensemble, as shown in panel (a) in Fig. \ref{fig6}.
Other features of spatiotemporal resource propagation remain qualitatively the same as shown in Fig. \ref{fig3} (c).

%

\widetext
\clearpage
\begin{center}
{\Large \textbf{Supplemental Material for\\[2pt]
\titleinfo}}\\[6pt]
Sreemayee Aditya,
Xhek Turkeshi,
Piotr Sierant
\end{center}

\setcounter{equation}{0}
\setcounter{figure}{0}
\setcounter{table}{0}
\setcounter{page}{1}
\renewcommand{\theequation}{S\arabic{equation}}
\setcounter{figure}{0}
\renewcommand{\thefigure}{S\arabic{figure}}
\renewcommand{\thepage}{S\arabic{page}}
\renewcommand{\thesection}{S\arabic{section}}
\renewcommand{\thetable}{S\arabic{table}}
\makeatletter

\renewcommand{\thesection}{\arabic{section}}
\renewcommand{\thesubsection}{\thesection.\arabic{subsection}}
\renewcommand{\thesubsubsection}{\thesubsection.\arabic{subsubsection}}

\vspace{1em}

\setcounter{section}{0}
\renewcommand{\thesection}{S\arabic{section}}
\renewcommand{\theequation}{S\arabic{equation}}
\renewcommand{\thefigure}{S\arabic{figure}}
\renewcommand{\thetable}{S\arabic{table}}
\setcounter{equation}{0}
\setcounter{figure}{0}
\setcounter{table}{0}

\section{Subsystem-size dependence of resource-monotone decay times}
As discussed in the Main Text, the resource monotones, under resource generating dynamics,  at late times relax exponentially toward the local maximally mixed state. Here, we quantify this decay using two complementary approaches. First, we fit the late-time profiles of resource monotones to exponentially decaying functions to extract the corresponding decay times and study their dependence on the subsystem size $L_{A}$. Second, we employ a threshold-based characterization of the decay time  by defining a threshold time $\tau_{\theta}$ such that for $t>\tau_\theta$ the resource measures fulfill $M(\rho_A(t)) < \theta \ll 1$, and we examine how $\tau_{\theta}$ scales with $L_{A}$.

Restricting to the first setup of the Main Text, i.e., resourceful dynamics intialized in an initially resourceless state, we
examine the late-time decay of the local resource content for three resource quantifiers: non-stabilizerness (a), coherence (b), and non-Gaussianity (c). Fitting the long time decays of monotones respectively with $\mathcal{L}(\rho_A(t)) \propto e^{-t/\tau^d_{\mathcal{L}}}$, $C_d(\rho_A(t)) \propto e^{-t/\tau^d_{c}}$, and $\mathcal{NG}(\rho_A(t)) \propto e^{-t/\tau^d_{N}}$, we extract the decay constants
$\tau^d_{R}$ (with $R=\mathcal{L}, c, N$).
We obtain $\tau^{d}_{\mathcal{L}}=(9.42,\,9.30,\,9.50)$ for $L_A=(2,\,3,\,4)$ (non-stabilizerness), $\tau^{d}_{c}=(9.90,\,10.01,\,9.65)$ for $L_A=(8,\,12,\,16)$ (coherence), and $\tau^{d}_{N}=(7.20,\,7.33,\,7.48,\,7.42,\,7.46)$ for $L_A=(8,\,16,\,24,\,32,\,36)$ (non-Gaussianity). Across all three diagnostics, the extracted $\tau^d_R$ values fluctuate by at most $\sim 5\%$ over the explored range of $L_A$, indicating that  to an essentially subsystem-size-independent late-time decay.

\begin{figure*}[h!]
\includegraphics[width=0.3\textwidth]{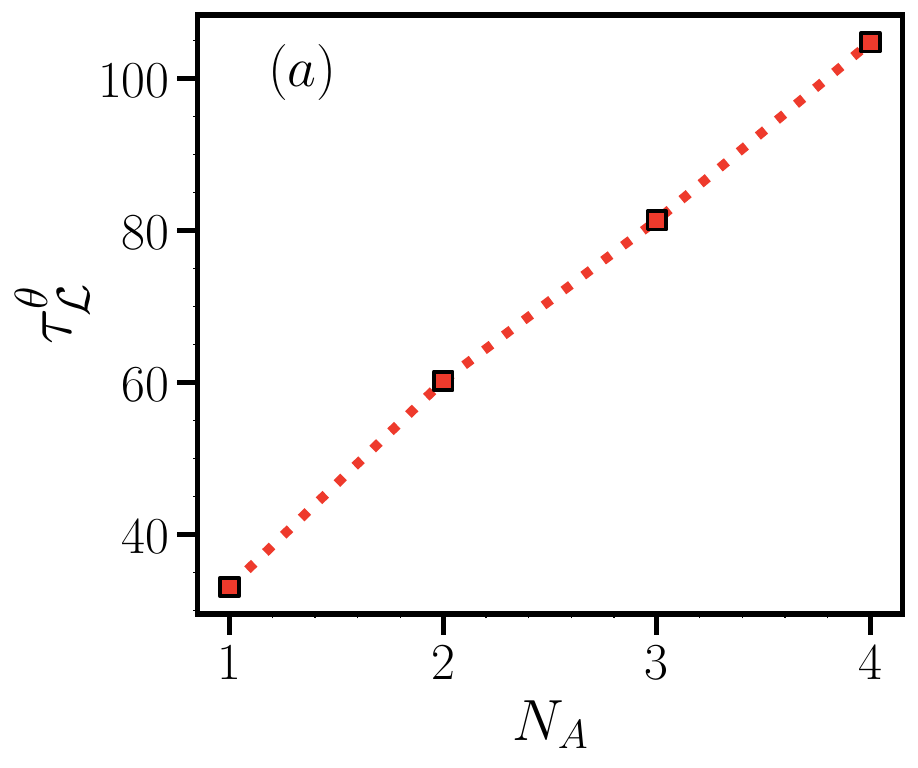}
\includegraphics[width=0.3\textwidth]{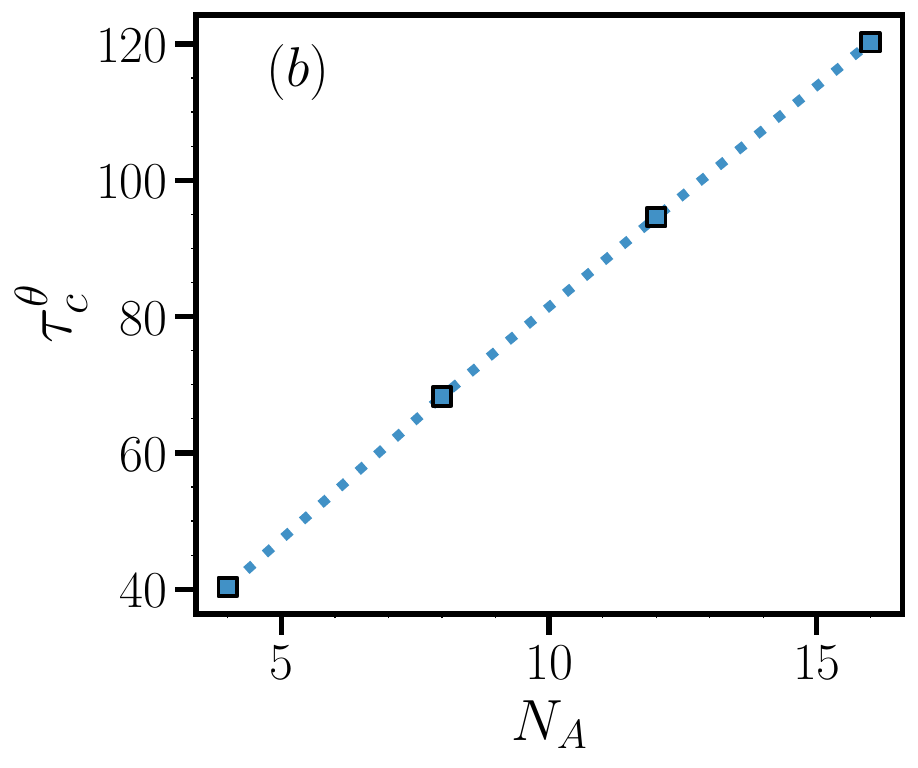}
\includegraphics[width=0.3\textwidth]{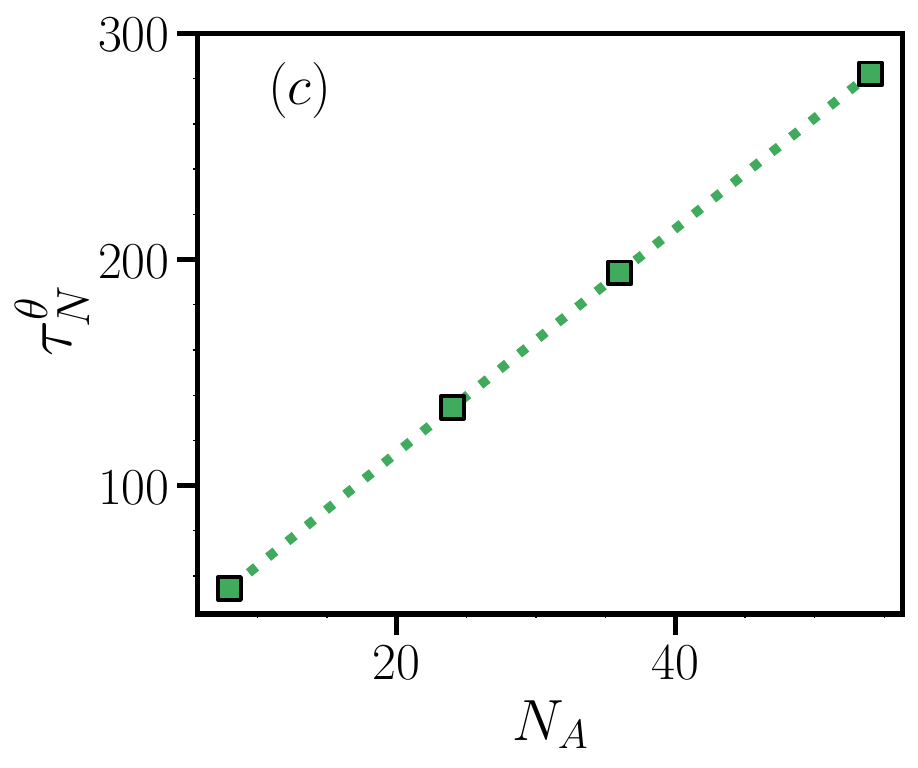}
\caption{Subsystem-size dependence of the threshold time $\tau_{\theta}$ for three representative resource measures: (a) non-stabilizerness ($L=24$ qubits), (b) coherence ($L=128$ qubits), and (c) non-Gaussianity ($L=128$ qubits). Starting from a resourceless initial state, the system evolves under resourceful dynamics generated by two-qubit operations for the setup shown in Fig. 2 in the main text. We define $\tau_{\theta}$ implicitly by the threshold condition $M(\tau_{\theta})=\theta$, with $\theta$ fixed to $0.01$ for nonstabilizerness and to $0.1$ for coherence and non-Gaussianity. Across all three measures, $\tau_{\theta}$ grows linearly with the subsystem size $L_A$.
}
\label{figs2:ttheta}
\end{figure*}

In Fig.~\ref{figs2:ttheta}, we turn to the threshold timescale $\tau_\theta$ for the same setup and the same three resources. Here the behavior is markedly different: $\tau_\theta$ increases linearly with $L_A$, indicating $\tau_\theta \propto L_A$.
Taken together, these observations point to a clear separation of dynamical regimes. The scaling $\tau_\theta \propto L_A$ is consistent with ballistic propagation of correlations, where the onset of resource depletion is set by a light-cone traversal across a subsystem of size $L_A$. By contrast, the subsequent late-time relaxation is governed by local dynamics, leading to an exponential decay of the resource monotones with an $L_A$-independent timescale $\tau_d$. In this sense, the ballistic spreading controls the \emph{onset} of resource depletion, while the \emph{rate} of the asymptotic decay is determined by a local equilibration process.

\end{document}